\documentclass{emulateapj}

\usepackage{txfonts}
\usepackage{xspace}
\usepackage{natbib}
\usepackage{graphicx}

\slugcomment{ Version of \today }

\shorttitle{Double-peaked outburst of A\,0535+26}
\shortauthors{Caballero et al.}

\begin{document}


\title{A double-peaked outburst of A\,0535+26 observed with \textsl{INTEGRAL}, \textsl{RXTE}, and \textsl{Suzaku}}


\author{I. Caballero\altaffilmark{1}, 
    K.~Pottschmidt\altaffilmark{2},
    D.~M.~Marcu\altaffilmark{2},
    L.~Barragan\altaffilmark{3},
    C.~Ferrigno\altaffilmark{4},
    D.~Klochkov\altaffilmark{5},
    J.~A.~Zurita Heras\altaffilmark{6},
    S.~Suchy\altaffilmark{5},
    J.~Wilms\altaffilmark{3},
    P.~Kretschmar\altaffilmark{7},
    A.~Santangelo\altaffilmark{5},
    I.~Kreykenbohm\altaffilmark{3},
    F.~F\"{u}rst\altaffilmark{8},
    R.~Rothschild\altaffilmark{9},
    R.~Staubert\altaffilmark{5},
    M.~H.~Finger\altaffilmark{10},
    A.~Camero-Arranz\altaffilmark{11},
    K.~Makishima\altaffilmark{12,13},
    T.~Enoto\altaffilmark{13},
    W.~Iwakiri\altaffilmark{14},
    Y.~Terada\altaffilmark{14}}

   \altaffiltext{1}{Laboratoire AIM, CEA/IRFU, CNRS/INSU, Universit\'e Paris Diderot, CEA DSM/IRFU/SAp, 91191 Gif-sur-Yvette, France} 
   \email{isabel.caballero@cea.fr}
   \altaffiltext{2}{CRESST, University of Maryland, Baltimore, MD 21250 /NASA GSFC, 
     Astrophysics Science Division, Code 661, Greenbelt, MD 20771, USA}
   \altaffiltext{3}{Dr.\,Karl Remeis-Sternwarte and ECAP, FAU Erlangen-Nuremberg, Sternwartstr. 7, 96049 Bamberg, Germany}
   \altaffiltext{4}{ISDC Data Centre for Astrophysics, 1290 Versoix, Switzerland}
   \altaffiltext{5}{Institut f\"ur Astronomie und Astrophysik, Sand 1, 72076 T\"ubingen, Germany}
   \altaffiltext{6}{Fran\c{c}ois Arago Centre, APC (UMR 7164 Universit\'{e} Paris Diderot, CNRS/IN2P3, 
     CEA/DSM, Observatoire de Paris), 13 rue Watt, 75205, Paris Cedex 13, France}
  \altaffiltext{7}{European Space Astronomy Centre (ESA/ESAC), Science Operations Department, Villanueva de la Ca\~{n}ada (Madrid), Spain}
   \altaffiltext{8}{Space Radiation Lab, California Institute of Technology, MC 290-17 Cahill, 1200 E. California Blvd., Pasadena, CA 91125, USA}
   \altaffiltext{9}{Center for Astrophysics and Space Science, UCSD,  La Jolla, CA, USA}
   \altaffiltext{10}{National Space Science and Technology Center, 320 Sparkman Drive NW, Huntsville, AL 35805 USA}
   \altaffiltext{11}{ Institut de Ci\`{e}ncies de l'Espai, (IEEC-CSIC), Campus UAB, Fac. de Ci\`{e}ncies, Torre C5, parell, 2a planta, 08193 Barcelona, Spain}
   \altaffiltext{12}{Department of Physics, The University of Tokyo, 7-3-1 Hongo, Bunkyo, Tokyo 113-0033, Japan}
   \altaffiltext{13}{Cosmic Radiation Laboratory, RIKEN, 2-1, Hirosawa, Wako City, Saitama 351-0198, Japan}
   \altaffiltext{14}{Graduate School of Science and Engineering, Saitama University, 255 Shimo-Okubo, Sakura, Saitama 338-8570, Japan}

 
\begin{abstract}
The Be/X-ray binary A\,0535+26 showed a normal (type I) outburst in August 2009. 
It is the fourth in a series of normal outbursts associated with the periastron, 
but is unusual by presenting a double-peaked light curve. The two peaks reached 
a flux of $\sim450$\, mCrab  in the 15--50\,keV range. 
We present results of the timing and spectral analysis 
of \textsl{INTEGRAL}, \textsl{RXTE}, and \textsl{Suzaku} observations of the outburst. 
The energy dependent pulse profiles and their evolution during the outburst are studied. 
No significant differences with respect to other normal outbursts are observed. The 
centroid energy of the fundamental cyclotron line shows no significant 
variation during the outburst. A spectral hardening with increasing luminosity is observed. We 
conclude that the source is accreting in the sub-critical regime. We discuss possible explanations for the double-peaked outburst. 
\end{abstract} 

\keywords{pulsars: individual (A\,0535+26) --- stars: magnetic field--- X-rays: binaries --- X-rays: stars}

\section{Introduction} 
The Be/X-ray binary A\,0535+26 was discovered during a giant outburst with \textsl{Ariel V}  
\citep{rosenberg75}. At a distance of  $d\sim2$\,kpc, it consists of 
the  B0~IIIe optical companion HDE\,245770 \citep{steele98} and a 
pulsating neutron star of spin period $P_{\mathrm{spin}}\sim103\,$s, in an 
eccentric orbit ($\mathrm{e}=0.47$) of period of $P_{\mathrm{orb}}=111.1$\,days 
\citep{finger06}.  The system presents different luminosity states associated with the activity of the Be star: 
quiescence, with X-ray luminosities $L_{\mathrm{X}}\lesssim10^{36}\,\mathrm{erg}\,\mathrm{s}^{-1}$ \citep{rothschild2012}, 
normal (type I) outbursts, with luminosities  $L_{\mathrm{X}}\sim10^{36-37}\,\mathrm{erg}\,\mathrm{s}^{-1}$, and giant (type II) outbursts, that can reach luminosities $L_{\mathrm{X}}>10^{37}\,\mathrm{erg}\,\mathrm{s}^{-1}$ (see, e.g., \citealt{finger96}). 
A\,0535+26 presents two cyclotron resonance scattering features (CRSFs or cyclotron lines) in its X-ray spectrum at 
$\sim$46 and $\sim$100\,keV \citep[see, e.g.,][and references therein]{caballero07}. Cyclotron lines are caused by resonant scattering of photons off electrons 
in the quantized Landau levels, and the cyclotron line energy is proportional to the magnetic field strength, 
$E_{\mathrm{cyc}}= \hbar eB/m_{\mathrm{e}}c=11.6\,\mathrm{keV}\cdot B/10^{12}\,\mathrm{G}$ \citep[e.g.,][]{schoenherr07}. 

Some accreting X-ray pulsars present a  negative correlation between the cyclotron line centroid 
energy and the X-ray luminosity (e.g., 4U\,0115+64\footnote{See, however, \cite{mueller12b}, where it is shown that for 4U\,0115+64 
the presence of a correlation depends on the model for the continuum.}, V\,0332+53, see \citealt{nakajima06,tsygankov06,mowlavi06}). 
Other sources, like Her\,X-1, GX\,304$-$1, and Swift\,J1626.6$-$5156 show the opposite trend (\citealt{staubert07},  \citealt{klochkov12}, \citealt{decesar12}). 
As discussed by \cite{becker12}, this bimodality is probably due to two different accretion 
regimes depending on the critical luminosity $L_{\mathrm{crit}}$. For sources with $L_{\mathrm{X}}>L_{\mathrm{crit}}$ 
(\emph{supercritical} sources), a radiative shock decelerates the infalling matter, with the 
height of the emission region increasing with increasing mass accretion rate,  
explaining the negative correlation between the cyclotron line energy and the X-ray luminosity. 
On the other hand, for sources with $L_{\mathrm{X}}<L_{\mathrm{crit}}$ (\emph{subcritical} sources), Coulomb interactions 
stop the infalling matter, and the height of the emission region decreases with increasing accretion rate, 
explaining the positive correlation observed (further details in \citealt{becker12}, see also 
\citealt{basko_sunyaev_76}, \citealt{staubert07}). 
Contrary to these sources, for A\,0535+26 no significant correlation between the cyclotron line energy and the X-ray 
luminosity has been observed \citep{terada06,caballero07}, with the exception of a flaring episode during the onset of a  
normal outburst observed in 2005 during which the cyclotron line energy significantly increased (see \citealt{caballero08a,postnov08}
for details). These results refer to pulse phase averaged spectroscopy; see however \cite{klochkov11} for pulse-to-pulse analysis in which 
a positive correlation between the flux and the cyclotron line energy was observed. 

\begin{figure}
  \plotone{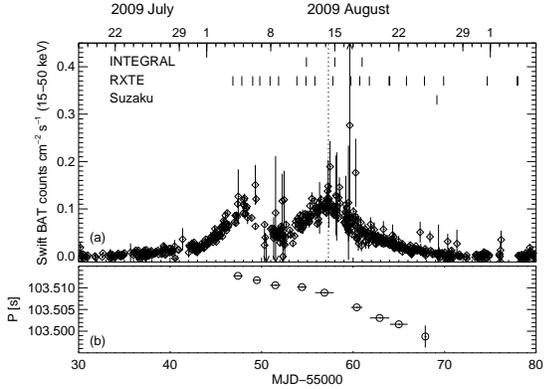} 
  \caption{(a) \textsl{Swift}/BAT orbital light curve (15--50\,keV) of the A\,0535+26 double-peaked 
    outburst. The tick marks indicate the times of the \textsl{INTEGRAL}, \textsl{RXTE}, and \textsl{Suzaku} 
    observations presented in this work. The vertical dotted line shows the time of periastron. (b) Pulse period 
    evolution derived from the \textsl{RXTE} observations.} 
  \label{fig:bat_lc}
\end{figure}

A\,0535+26 showed a giant outburst and a series of normal ones in 2009, all of them 
associated with the periastron passage of the neutron star \citep{caballero09_b}. 
The giant outburst reached a flux of 
$F_{\mathrm{(15-50)\,keV}}=(1.17\pm0.04)\,\mathrm{counts}\,\mathrm{s}^{-1}\,\mathrm{cm}^{-2}$
($\sim5.3$\,Crab\footnote{Fluxes derived from the daily \textsl{Swift}/BAT transient monitor results provided by the Swift/BAT team. 
$\sim$0.22 counts\,s$^{-1}$\,cm$^{-2}$ correspond to $1\,$Crab.}) at MJD~55179.0 and lasted about 40 days.  
In this work we focus on a double-peaked outburst that preceded the giant one. We present \textsl{INTEGRAL}, \textsl{RXTE}, and \textsl{Suzaku} 
observations of the outburst. The  \textsl{Swift}/BAT light curve of the outburst and the times of the 
observations are shown in Fig.~\ref{fig:bat_lc} (a). The first peak of the outburst reached a flux of 
$F_{\mathrm{(15-50)\,keV}}=(0.100\pm0.004)\,\mathrm{counts}\,\mathrm{s}^{-1}\,\mathrm{cm}^{-2}$ ($\sim$\,455\,mCrab$^{16}$) at MJD 55048.0, 
that decreased to $F_{\mathrm{(15-50)\,keV}}=(0.047\pm0.002)$\,counts\,s$^{-1}$\,cm$^{-2}$ ($\sim$214\,mCrab)  at MJD 55051.0 and rose 
again, reaching $F_{\mathrm{(15-50)\,keV}}=(0.104\pm0.005)$\,counts\,s$^{-1}$\,cm$^{-2}$ ($\sim$471\,mCrab) at MJD 55058.0 around periastron. 
The observations are described in Sec.~\ref{sec:obs}.  The results of the timing and spectral analyses are 
presented in Sec.~\ref{sec:results}, and the results are discussed in Sec.~\ref{sec:concl}. 

\section{Observations and data analysis}\label{sec:obs}

We made use of \textsl{INTEGRAL} data collected with the gamma ray spectrometer 
SPI \citep[20\,keV--8\,MeV,][]{vedrenne03}, the imager IBIS   
\citep[15~keV--10~MeV,][]{ubertini03}, and the X-ray monitor 
JEM-X \citep[3--35\,keV,][]{lund03}. Three pointed observations were performed around the 
second peak of the outburst (MJD start 55054.89, 55057.99, and 55060.99). 
The JEM-X and SPI data analyses were performed using the standard analysis package OSA v9. For IBIS, we used 
OSA v10, that contains a new energy calibration for ISGRI. 

\textsl{RXTE} \citep{bradt93} performed regular pointed observations of A\,0535+26 during the outburst between MJD 55046.86--55078.00 
(Proposal ID P94323). We used data from the Proportional Counter Array PCA \citep[2--60keV,][]{jahoda96} and the
High Energy X-ray Timing Experiment HEXTE \citep[20--200\,keV,][]{rothschild98}. 
The \textsl{RXTE} data were analyzed using HEASOFT v6.7. We restricted the PCA spectral analysis to energies
above 5\,keV due to a feature around 5$\,$keV caused by instrumental Xe L edges \citep{rothschild06}. 

\textsl{Suzaku} performed one observation of A\,0535+26 during the decay of the outburst (MJD start 55067.96, ObsID 404054010). 
We used data from the two main instruments, the X-ray Imaging Spectrometer \citep[XIS,][]{koyama07} and the Hard 
X-ray Detector \citep[HXD,][]{takahashi07}. 
The data analysis was performed using HEASOFT v6.12 and CALDB versions 20110913 for HXD, 
20120209 for XIS, and 20110630 for XRT. We used XIS data between 0.5--10 keV for XIS 1, and data between 
0.7--10 keV for XIS 0, 3. XIS data were binned as in \cite{nowak11}. 
Because of the XIS calibration uncertainties, especially around the Si K edge at $\sim$1.8\,keV, a systematic error of 1\% was 
assumed for the XIS data. 
The spectral analysis was performed with \textsl{XSPEC} v12.7.0.   

\section{Results}\label{sec:results}

\subsection{Timing analysis}\label{sec:timing}
\begin{figure}
  \plotone{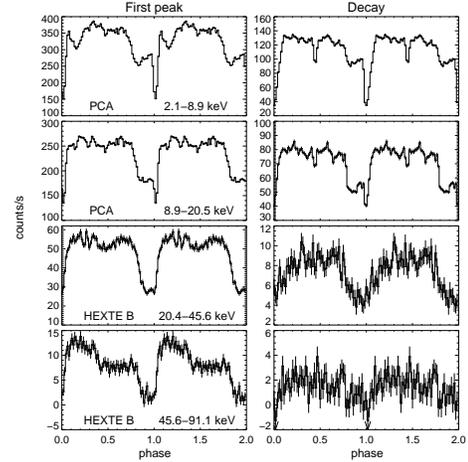}
  \caption{\textsl{RXTE} PCA and HEXTE energy dependent pulse profiles for one observation around the first peak 
    (\textsl{left}, MJD 55047.86, ObsID 94323-02-03-01) and one observation during the decay of the outburst 
    (\textsl{right}, MJD 55065.84, ObsID 94323-03-02-01). The energy ranges are indicated in 
    the left panels, and are the same for both observations. The PCA count rate is given in counts\,$\mathrm{s}^{-1}\,$PCU$^{-1}$.
    Two pulse cycles are shown for clarity.
  }\label{fig:pp}
\end{figure} 

\begin{figure*}
\includegraphics[angle=90,width=0.48\textwidth]{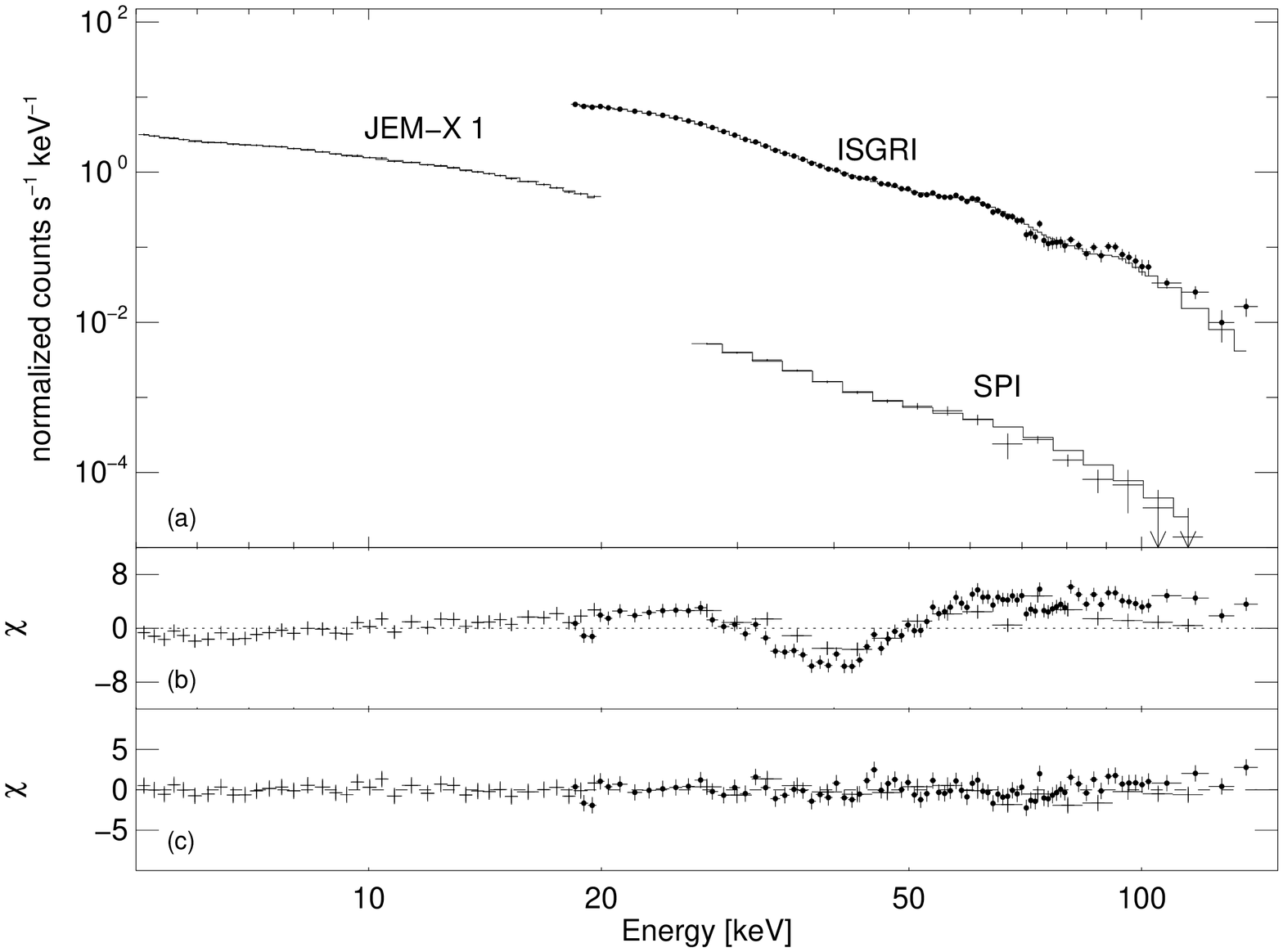}
\includegraphics[angle=90,width=0.48\textwidth]{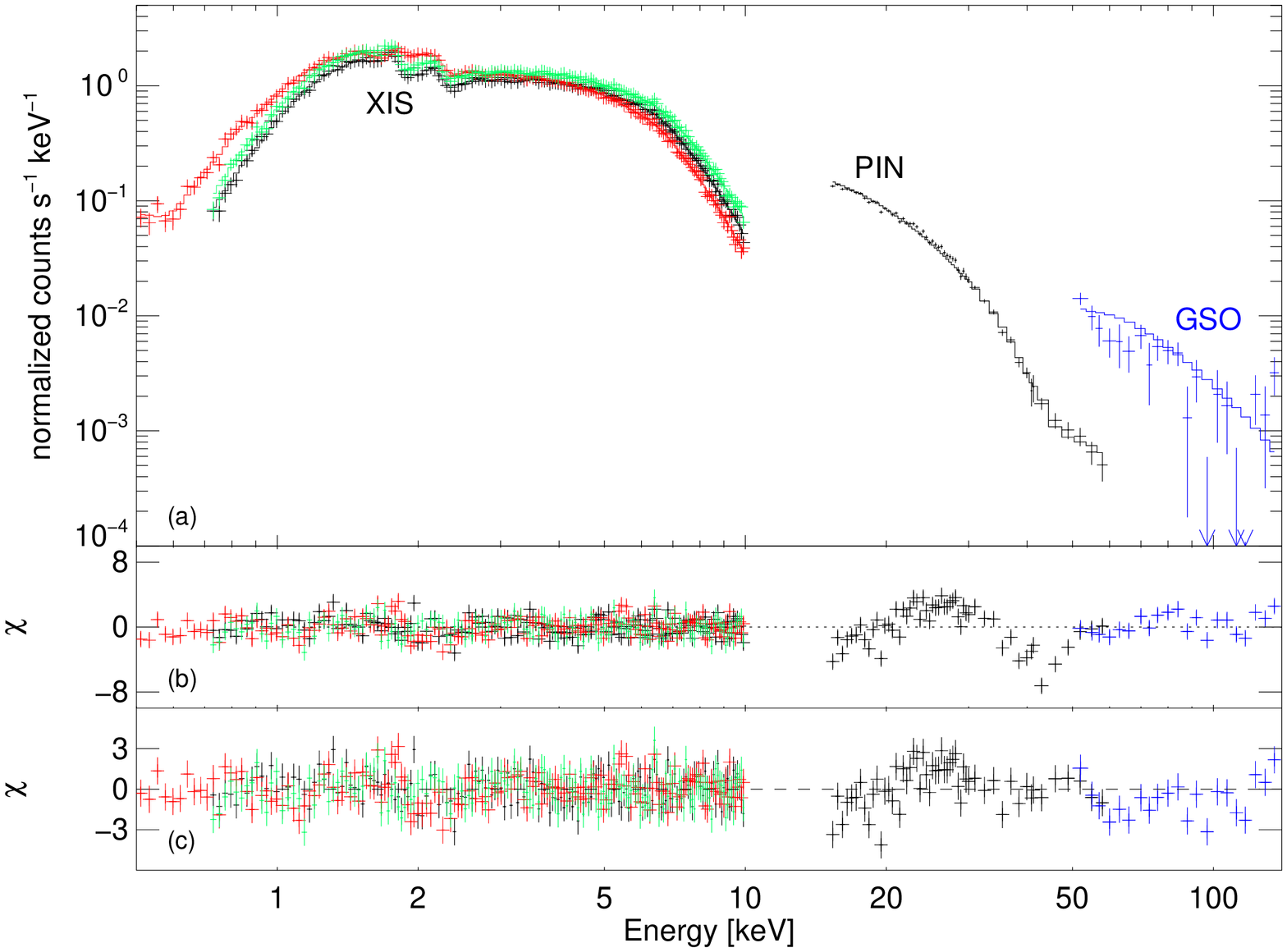}
 \caption{\label{fig:spec}
    \textsl{Left:} \textsl{INTEGRAL} 
    spectrum of A\,0535+26 (MJD 55057.99). (a) Data and 
    best fit model. (b) Residuals of a fit including no absorption lines in the model. (c) Residuals 
    of a fit including one line at $\sim46$\,keV in the model. The filled circles represent the ISGRI data.  
    The average flux during the observation 
    $F_{20-100\,\mathrm{keV}}\sim5.0\times10^{-9}\,\mathrm{erg\,cm^{-2}\,s^{-1}}$. 
   \textsl{Right:} \textsl{Suzaku} spectrum of A\,0535+26 (MJD 55067.96). (a), (b), (c) 
    have the same meaning as in the left panel. The black, red, and green symbols 
    represent XIS 0, 1, and 3 data respectively. The average flux during the observation  
   $F_{20-100\,\mathrm{keV}}\sim0.5\times10^{-9}\,\mathrm{erg\,cm^{-2}\,s^{-1}}$.} 
\end{figure*} 

We studied the pulse period evolution using the phase-connection technique described by \cite{staubert09}. 
We extracted PCA light curves with 15.6\,ms resolution in the 3.3--42.8\,keV range. Barycentric and orbital corrections
were applied to the light curves, using the ephemeris from \cite{finger06}. 
The evolution of the pulse period is shown in Fig.~\ref{fig:bat_lc} (b). A decrease in the pulse period along the 
outburst is observed. A linear fit can be used to describe the pulse period evolution between MJD 55046.89--55057.88, 
during the two main peaks of the outburst. We find a pulse period of $P=103.51263\pm 0.00004\,$s and spin-up of 
$\dot{P}=(-0.459\pm0.005)\times10^{-8}$\,s\,s$^{-1}$ at MJD 55047.4. 
The pulse period during the decay of the outburst (MJD $>$ 55057.88) can also be described with a linear fit. 
We obtain a spin period of $P=103.50886\pm 0.00001\,$s for MJD 55056.87 and a spin-up of 
$\dot{P}=(-1.104\pm0.002)\times10^{-8}$\,s\,s$^{-1}$. 

Making use of these pulse period values to fold the light curves, we studied the evolution of the energy 
dependent pulse profiles using all the \textsl{RXTE} pointed observations. 
During the two main peaks of the outburst, the pulse profiles are remarkably stable, showing a complex structure 
at low energies, and becoming simpler at higher energies. The shape of the pulse profiles at low energies 
slightly varies towards the decay of the outburst. 
As an example, the pulse profiles for one observation during the first peak of the outburst and one observation 
during the decay are shown in Fig.~\ref{fig:pp}. 

\subsection{Spectral analysis:}\label{sec:spec}   

We studied the broad band spectral continuum and its evolution during the outburst with \textsl{RXTE}, \textsl{INTEGRAL}, 
and \textsl{Suzaku}. It can be described by a powerlaw with an exponential cutoff, typical for accreting X-ray binaries (see, e.g., 
\citealt{white83}). We used the  \textsl{XSPEC} model  {\tt{cutoffpl}}, given by 
$F(E)\propto E^{-\Gamma}e^{-E/E_{\mathrm{fold}}}$, where $\Gamma$ and $E_{\mathrm{fold}}$ are the photon 
index and folding energy respectively. 

The \textsl{Suzaku} spectrum was modeled including an additional black body component, photoelectric 
absorption and a Gaussian line in emission to account for the Fe K$\alpha$ line. 
The presence of the Fe K$\alpha$ line is marginal, and the line energy and width have been fixed to 6.4 and 0.5 keV 
respectively. Its inclusion in the model yields an improvement in the $\chi^{2}_{\mathrm{red}}$/d.o.f from 
1.43/581 to 1.37/580. The equivalent width of the line is EW (Fe K$\alpha$)=$70\pm2$\,eV. The inclusion of the black body 
component in the model leads to an improvement  in the  $\chi^{2}_{\mathrm{red}}$/d.o.f. 
from 3.01/582 to 1.37/580, giving an F-test significance $>99.99\%$. The best fit value of the black body temperature 
is $k_{B}T=1.26^{+0.05}_{-0.04}$\,keV. 
Setting the cross-sections and the abundances to those from \cite{verner96} and \cite{wilms00} respectively, 
we obtained an absorption column of $N_{\mathrm{H}}=0.70\pm0.03\times10^{22}$\,atoms\,cm$^{-2}$. 

To model the \textsl{RXTE} spectra, we added a Gaussian emission line to account for residuals present in 
the 5--7\,keV range. The energy of the line was left free, and the width of the line was frozen to the best fit value
in each observation. 
The line energy obtained for the different observations varies between 
5.0 and 6.6\,keV, consistent with the instrumental Xe L edge at $\sim5\,$keV and the 6.4 keV Fe K$\alpha$ line. 
It was not possible to model the two components independently. 

In the \textsl{INTEGRAL}, \textsl{Suzaku}, and most of the \textsl{RXTE} observations,  
a significant absorption-like feature is present in the residuals at $E\sim46$\,keV, 
that we model using a Gaussian optical depth profile $\tau(E)=\tau e^{-(E-E_{\mathrm{cyc}})^{2}/(2\sigma^{2})}$, which modifies the continuum 
as $F'(E)=F(E)e^{-\tau(E)}$. As an example, the broad band spectrum of the second \textsl{INTEGRAL} observation is shown in Fig.~\ref{fig:spec} (left).  
In this case, the inclusion of a cyclotron line in the model yields an improvement in the $\chi^{2}_{\mathrm{red}}$/d.o.f.  
from 8.8/132 to 0.88/129. The \textsl{Suzaku} broad band spectrum is shown in Fig.~\ref{fig:spec} (right). 
The inclusion of a cyclotron line in this case improves the $\chi^{2}_{\mathrm{red}}$/d.o.f.  from 1.73/583 to 1.37/580. 

\begin{figure} 
\includegraphics[width=0.48\textwidth]{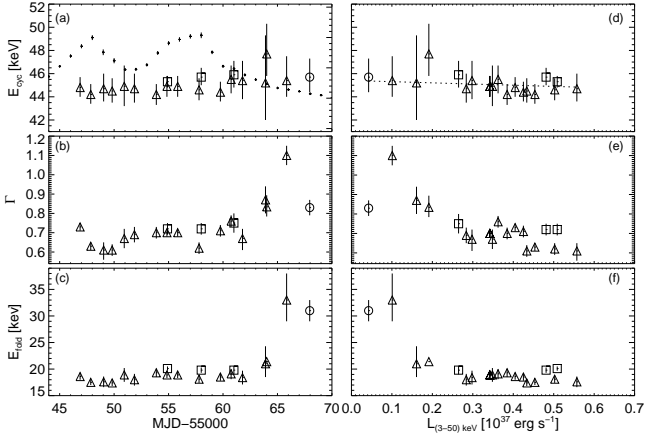}
  \caption{\textsl{Left}: Evolution of the cyclotron line centroid energy $E_{\mathrm{cyc}}$ (a), photon index 
    $\Gamma$ (b), and folding energy $E_{\mathrm{fold}}$ (c) with time.  
    In panel (a), the crosses represent the \textsl{Swift}-BAT light curve in arbitrary units. 
    \textsl{Right:} evolution of $E_{\mathrm{cyc}}$ (d), $\Gamma$ (e), 
    and $E_{\mathrm{fold}}$ (f) with the (3--50)\,keV X-ray luminosity. 
    The \textsl{RXTE}, \textsl{INTEGRAL}, and \textsl{Suzaku} observations  are represented with triangles, squares, and a circle respectively. 
   In panel (d), the dashed line represents a linear fit to the data. Error bars are at 90\% confidence. }\label{fig:Ecyc_cont}
\end{figure}

We studied the evolution of the cyclotron line centroid energy, photon index, and folding energy as a function of time and 
X-ray luminosity. To study the evolution of the different spectral parameters, we selected the first 16 \textsl{RXTE} observations (MJD 55046.86--55065.84), 
which are the ones where the cyclotron line was significantly detected. The first harmonic line at $E\sim100\,$keV, observed in brighter outbursts of the source 
(e.g., \citealt{caballero07}), is not significantly detected in the observations presented here. 
The results are shown in Fig.~\ref{fig:Ecyc_cont}. No dependency of the cyclotron line energy 
with the X-ray luminosity is found. We checked that the choice of continuum model does not affect the line 
parameters by fitting the continuum using an exponentially cutoff power
law, a Fermi Dirac cutoff power law \citep{tanaka86}, and a power law with an exponential
cutoff starting at a cutoff energy (the {\tt{highecut}} model of \textsl{XSPEC}). In
all three cases the line parameters remain the same within their error
bars  (see also \citealt{caballero09_c}).  A linear fit to the $E_{\mathrm{cyc}}$ vs. $L_{\mathrm{X}}$ values gives a slope of 
$(-3\pm3)\,$keV$/10^{37}\mathrm{erg}\,\mathrm{s}^{-1}$ using \textsl{RXTE} data only, and 
$(-0.9\pm2.0)$\,keV/$10^{37}\mathrm{erg}\,\mathrm{s}^{-1}$ using \textsl{RXTE}, \textsl{INTEGRAL}, and \textsl{Suzaku} data.
We included inter-calibration constants for the different instruments in the separate \textsl{RXTE}, \textsl{INTEGRAL}, and \textsl{Suzaku} spectral analysis. 
For the \textsl{Suzaku} analysis, we fixed the constant to 1 for XIS 0 and allowed the others to vary. 
In the case of the GSO detector of the HXD, we fixed the constant to $c_{\mathrm{GSO}}=0.88$ (the value obtained before adding 
the cyclotron line to the model), because when unconstrained, the best fit results in an unrealistic value of 
$c_{\mathrm{GSO}}=0.4\pm0.1$. By doing this, the folding energy decreases from 
$E_\mathrm{fold}=44^{+9}_{-6}\,$ to $31\pm2\,$keV, while the powerlaw index and the cyclotron line energy remain constant within the errors.
As seen in the middle and bottom panels of Fig.~\ref{fig:Ecyc_cont}, the photon index significantly decreases with 
increasing luminosity. The folding energy remains constant during the main part of the outburst, down to a luminosity of 
$L_{(3-50)\,\mathrm{keV}}\sim0.26\times10^{37}\mathrm{erg}\,\mathrm{s}^{-1}$, and shows an increase in the last 
observations of the outburst. Both the folding energy and the photon index increase with decreasing luminosities.  
The values plotted in Fig.~\ref{fig:Ecyc_cont} have been obtained leaving these two parameters free in the spectral fits. 
The increase of the photon index with decreasing X-ray luminosity indicates a softening of the spectrum, while 
the increase of the folding energy with decreasing X-ray luminosity indicates a hardening of the spectrum. 
In order to check if the spectrum becomes softer or harder as the luminosity decreases, 
in the last three \textsl{RXTE} observations we fixed the folding energy to the mean value of the first 13 observations, $E_{\mathrm{fold}}=18.4\,$keV. 
By doing this, a slope of $(-0.44\pm0.04) /10^{37}\mathrm{erg}\,\mathrm{s}^{-1}$ is obtained 
from a linear fit to the $\Gamma$ vs. $L_{\mathrm{X}}$ values,  
with a Pearson correlation coefficient of 0.82. This result indicates a softening of the spectrum with decreasing luminosity. 

\section{Discussion}\label{sec:concl} 

We have presented the first observations of a double-peaked outburst of A\,0535+26 with \textsl{RXTE}, 
\textsl{INTEGRAL}, and \textsl{Suzaku}. The double-peaked outburst was the precursor of a giant one in December 2009. 
Two double-peaked outbursts were also observed before a giant one in 1994 \citep{finger96}, while other giant outbursts do not show any precursor outbursts 
\citep{tueller05}. Another outburst with a peculiar shape (two separate peaks) was observed two orbital phases after the 2009 giant outburst 
\citep[see][]{camero12}. 

We measured a pulse period of  $P=103.51263\pm 0.00004\,$s and spin-up of 
$\dot{P}=(-0.459\pm0.005)\times10^{-8}$\,s\,s$^{-1}$ at MJD 55047.4, and a spin-up of 
$\dot{P}=(-1.104\pm0.002)\times10^{-8}$\,s\,s$^{-1}$ at MJD 55056.87. These results are in agreement with the values obtained 
with \textsl{Fermi}-GBM by \cite{camero12}, who also present the long-term pulse period history of A\,0535+26.  
We note that the changes in the spin-up rate along the outburst could be due to the outdated 
orbital solution. A new orbital solution is required to explore the pulse period evolution in more detail. 
A spin-up during normal outbursts has also been observed in the past (\citealt{caballero08a}), providing evidence for an accretion disk around the neutron star 
during normal outbursts. The energy and luminosity dependent pulse profiles are remarkably stable compared to past observations of the source 
(see, e.g., \citealt{caballero07} and \citealt{naik08}), and consistent with those obtained with \textsl{Fermi}-GBM by \citet{camero12} during the same period.

We observe no significant variation of the cyclotron line energy with the X-ray luminosity during the double-peaked 
outburst, and a significant decrease of the photon index $\Gamma$ (spectral hardening) with increasing luminosity. A similar 
behavior has been observed in other outbursts of the source (e.g., \citealt{caballero08a,mueller_d_12}). 
In accreting X-ray pulsars, a positive or negative correlation of $E_{\mathrm{cyc}}$ with $L_{\mathrm{X}}$ 
is related to different accretion regimes. 
Following the theoretical predictions from \cite{becker12}, the critical luminosity for A\,0535+26 corresponds to 
$L_{\mathrm{crit}}\sim6.78\times10^{37}\,$erg\,s$^{-1}$.  
The observations presented here are well below that limit, with $L_{(3-50)\,\mathrm{keV}}$ ranging between $\sim0.04$ and $\sim0.56\times10^{37}\,$erg\,s$^{-1}$, 
and therefore A\,0535+26 is probably accreting in the sub-critical regime during the outburst. A positive correlation between $E_{\mathrm{cyc}}$ and $L_{\mathrm{X}}$ 
is expected for sub-critical sources. The fact that no correlation is observed might be due to the fact that 
our observations are near or below the Coulomb stopping limit, $L_{\mathrm{coul}}$, where little variation of $E_{\mathrm{cyc}}$ with 
$L_{\mathrm{X}}$ is expected \citep[see][for further details]{becker12}. 

The double-peaked shape of the light curve could be due to perturbations in the Be disk around the optical companion. 
\cite{moritani11} showed that A\,0535+26 exhibited strong H$\alpha$ variability during, before, and after the giant outburst in 2009, 
and suggested that strong perturbations in the Be disk started about one cycle before the giant outburst. 
In addition, the H$\alpha$ EW and V-band brightness showed an anti-correlation before the 2009 giant outburst \citep{yan12,camero12}.  
This indicates that a mass ejection event took place before the giant outburst, producing a low-density region in the inner part of the disk that could 
explain the double-peaked profile \citep{yan12}.  Interestingly, \citet{yan12} reported a similar H$\alpha$ EW and V-band brightness evolution 
before the 1994 outburst, when two double-peaked outbursts took place before the giant one \citep{finger96}. 
Such an anticorrelation was not observed before the 2005 giant outburst, that took place without precursor outbursts. 
Double-peaked light curves have recently been observed in other sources, for instance GX\,304$-$1 \citep{nakajima12} and XTE J\,1946+274 
\citep{mueller12a}. While the spectral shape again remained relatively constant over the outburst for the latter, this is a special case, 
since two outbursts per orbit are observed. In the case of GX\,304$-$1 the double-peaked outburst took place before a giant one, similar to 
what has been observed for A\,0535+26, suggesting that double-peaked outbursts could be indicators of upcoming giant outbursts \citep{nakajima12}. 
The double-peaked outburst in the case of GX\,304$-$1 was similar to the one of A\,0535+26 in terms of duration, lasting about 40 days. 
While in the case of A\,0535+26 the two peaks reached a similar flux level, in the case of GX\,304$-$1 the second peak reached a flux two times higher 
than the first one. The intensities of the two peaks were  $F_{\mathrm{(15-50)\,keV}}=(0.098\pm0.004)$\,counts\,s$^{-1}$\,cm$^{-2}$ ($\sim$445\,mCrab) at  
MJD 56075 and $F_{\mathrm{(15-50)\,keV}}=(0.196\pm0.008)$\,counts\,s$^{-1}$\,cm$^{-2}$ ($\sim$891\,mCrab) at MJD 56087, with 
a flux between the two peaks that dropped down to  $F_{\mathrm{(15-50)\,keV}}=(0.021\pm0.001)$\,counts\,s$^{-1}$\,cm$^{-2}$ ($\sim$95\,mCrab) 
at MJD 56082. Note that the distances of the two sources are rather similar, with GX\,304$-$1 being at $\sim2.4$\,kpc \citep{parkes80}. 
The giant outburst of GX\,304$-$1 lasted about 50 days, showing a first weaker peak that reached a X-ray flux of 
$F_{\mathrm{(15-50)\,keV}}=(0.163 \pm0.007)$\,counts\,s$^{-1}$\,cm$^{-2}$ ($\sim$741\,mCrab) 
 at MJD 56208. The flux then dropped down to zero (formally $F_{\mathrm{(15-50)\,keV}}=(0.004\pm0.021)$\,counts\,s$^{-1}$\,cm$^{-2}$ at MJD 56223), 
and increased again, reaching $F_{\mathrm{(15-50)\,keV}}=(0.406 \pm 0.015)$\,counts\,s$^{-1}$\,cm$^{-2}$ ($\sim$ 1.85\,Crab) at MJD 56234.
Further simultaneous X-ray and optical observations of double-peaked outbursts are needed to better understand their origin and relation to giant ones.


\begin{acknowledgements}
We thank the anonymous referee for his/her useful comments, 
the \textsl{RXTE}, \textsl{INTEGRAL}, and \textsl{Suzaku} teams for the scheduling of the observations, and 
ISSI (Bern) for their hospitality during our collaboration meetings.  
IC thanks Philippe Laurent for the help with the \textsl{INTEGRAL} analysis, Yuuki Moritani for useful discussions, and 
acknowledges financial support from the French Space Agency CNES through CNRS. KP and DMM acknowledge support from 
NASA guest observer grants NNXIOAJ47G for \textsl{INTEGRAL} cycle 6 and  NNXIOAJ48G for \textsl{Suzaku} cycle 4. 
JW and IK acknowledge partial funding from the Deutsches Zentrum f\"ur Luft- und Raumfahrt under contract number 50 OR 1113.
ACA is supported by the grants AYA2009-07391 and SGR2009-811, as well as the Formosa program TW2010005 and iLINK program 2011-0303.
\end{acknowledgements}


\end{document}